\documentclass[sigconf, nonacm, screen]{acmart}

\usepackage[noend]{algpseudocode}
\usepackage{float}
\usepackage{subcaption}
\usepackage{tabularx}
\usepackage{booktabs}
\usepackage{algorithm2e}
\usepackage{multirow}
\usepackage{booktabs}
\usepackage{tablefootnote}
\usepackage{threeparttable}

\DeclareMathOperator*{\argmax}{arg\,max}

\usepackage{amsmath,amssymb}
\usepackage{pifont}
\usepackage{xcolor}
\usepackage{listings}

\usepackage{fancyhdr}

\definecolor{codegreen}{rgb}{0,0.6,0}
\definecolor{codegray}{rgb}{0.5,0.5,0.5}

\definecolor{backcolour}{RGB}{245,248,250}
\definecolor{emph}{RGB}{166,88,53}
\definecolor{nightblue}{RGB}{9,49,105}
\definecolor{keywords}{RGB}{207,33,46}
\definecolor{lightpurple}{RGB}{130,81,223}

\lstdefinestyle{mystyle}{
    backgroundcolor=\color{backcolour},   
    commentstyle=\color{codegreen},
    keywordstyle=\color{keywords},
    stringstyle=\color{nightblue},
    basicstyle=\ttfamily\footnotesize,
    breakatwhitespace=false,         
    breaklines=true,                 
    captionpos=b,                    
    keepspaces=true,                 
    numberstyle=\small\color{codegray},
    numbers=left,                    
    numbersep=5pt,   
    xleftmargin=0.2cm,
    aboveskip=0.2cm,
    belowskip=0.1cm,
    showspaces=false,                
    showstringspaces=false,
    showtabs=false,                  
    tabsize=2,
    frame=shadowbox,
    emph={sem_filter,sem_join,sem_topk,sem_sim_join,sem_agg,sem_map,sem_extract,sem_cluster_by,sem_search,sem_index,load_sem_index,sem_partition_by,sem_group_by},
    emphstyle={\color{lightpurple}},
}

\newif\ifcomments
\commentstrue
\ifcomments

\else
    \providecommand{\pb}[1]{}
\fi

\newif\ifgap
\gaptrue
\ifgap
    \providecommand{\gap}[1]{{\color{purple}{/* gap: #1 */}}}
    
\else
    \providecommand{\gap}[1]{}
\fi

\newif\ifrev
\ifrev

\else

\fi

\newcommand{\heading}[1] {{\noindent{\textbf{\emph{#1}}} }}

\newcommand{\headingtwo}[1] {{\noindent{\underline{#1}}} }

\newcommand{\comma}[1] {{\textit{,\space\space\space}{#1} }}

\newcommand{\sys}{LOTUS\xspace}

\definecolor{logocolor}{RGB}{138, 72, 213}                %

\begin{document}

\fancyhead[L]{\href{https://lotus-data.github.io}{\raisebox{-0.1\height}{\includegraphics[height=1.5em]{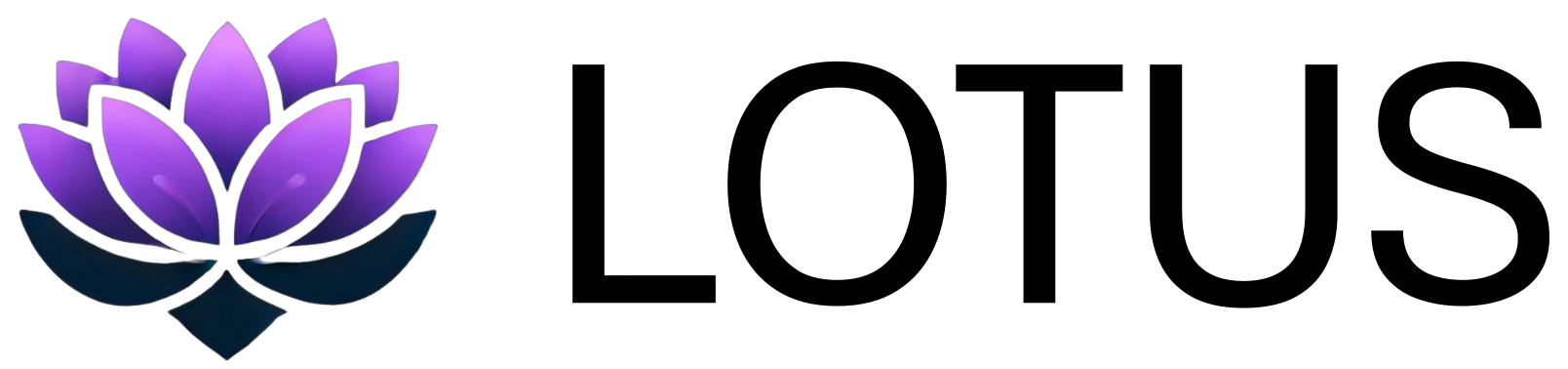}}}}
\fancyhead[C]{\href{https://github.com/lotus-data/lotus}{\raisebox{-0.1\height}{\includegraphics[height=1em] {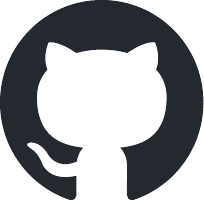}}}}
\fancyhead[R]{Semantic Operators: Declarative AI-based Data Processing}
\setlength{\headheight}{15pt}
\pagestyle{fancy}

\renewcommand{\headrulewidth}{0.4pt}

\fancyfoot[C]{\thepage \hspace{1pt}}

\title{Semantic Operators: A Declarative Model for Rich, AI-based Data Processing}


\author{Liana Patel$^\dagger$,  Siddharth Jha$^\ddagger$, Melissa Pan$^\ddagger$, Harshit Gupta$^\dagger$, Parth Asawa$^\ddagger$,\\ Carlos Guestrin$^\dagger$, Matei Zaharia$^\ddagger$}
\affiliation{%
  \institution{$\dagger$ Stanford University, $\ddagger$ UC Berkeley}
  \institution{Correspondence: lianapat@stanford.edu}
  \institution{
  \begin{center}
  \href{https://github.com/lotus-data/lotus}{\raisebox{-0.1\height}{\includegraphics[height=1em]
{figures/github-mark.pdf}}
{LOTUS Repository and Tutorials}}\end{center}
  }
}

\renewcommand{\shortauthors}{Patel et al.}

\begin{abstract}
The semantic capabilities of large language models (LLMs) have the potential to enable rich analytics and reasoning over vast knowledge corpora. 
Unfortunately, existing systems either empirically optimize expensive LLM-powered operations with \emph{no performance guarantees}, or serve a limited set of \emph{row-wise} LLM operations, providing limited robustness, expressiveness and usability.
We introduce \emph{semantic operators}, the first formalism for declarative and general-purpose AI-based transformations based on natural language specifications (e.g., filtering, sorting, joining or aggregating records using natural language criteria).  Each operator opens a rich space for execution plans, similar to relational operators. Our model specifies the expected behavior of each operator with a high-quality gold algorithm, and we develop an optimization framework that reduces cost, while providing accuracy guarantees with respect to a gold algorithm.
Using this approach, we propose several novel optimizations to accelerate semantic filtering, joining, group-by and top-k operations by up to $1,000\times$.  We implement semantic operators in the \sys system and demonstrate \sys' effectiveness on real, bulk-semantic processing applications, including fact-checking, biomedical multi-label classification, search, and topic analysis. We show that the semantic operator model is expressive, capturing state-of-the-art AI pipelines in a few operator calls, and making it easy to express new pipelines that match or exceed quality of recent LLM-based analytic systems by up to $170\%$, while offering accuracy guarantees.
Overall, \sys programs match or exceed the accuracy of state-of-the-art AI pipelines for each task while running up to $3.6\times$ faster than the highest-quality baselines. 
    \sys is publicly available at \url{https://github.com/lotus-data/lotus}.
\end{abstract}

\maketitle

\section{Introduction}
The powerful semantic capabilities of modern  language models (LLMs) create exciting opportunities for building AI-based analytics systems that reason over vast knowledge corpora. Many applications require complex reasoning over large amounts of data, including both unstructured and structured data. For example a researcher reviewing recent ArXiv~\cite{arxivorg} pre\-prints may want to quickly obtain a summary of relevant papers from the past week, or find the papers that report the best performance for a particular task and dataset. Similarly, a medical professional may automatically extract biomedical characteristics and candidate diagnoses from many patient reports~\cite{doosterlinck_biodex_2023}. Likewise, organizations wish to automatically digest lengthy transcripts from internal meeting transcripts and chat histories to validate hypotheses about their business~\cite{discovery}.

Each of these tasks require a form of \emph{bulk semantic processing}, where the analytics system must process large amounts of data and apply semantic-based analysis 
across a whole dataset. Supporting the full generality of these applications with efficient and easy-to-use analytics systems would have a transformative impact, similar to what RDBMSes had for tabular data. This prospect, however, raises two challenging questions: first, \emph{how should developers express semantic queries}, and secondly, \emph{how should we design the underlying analytics system to achieve high efficiency and accuracy}.

\begin{figure}[!t]
  \centering
  \includegraphics[width=\linewidth]{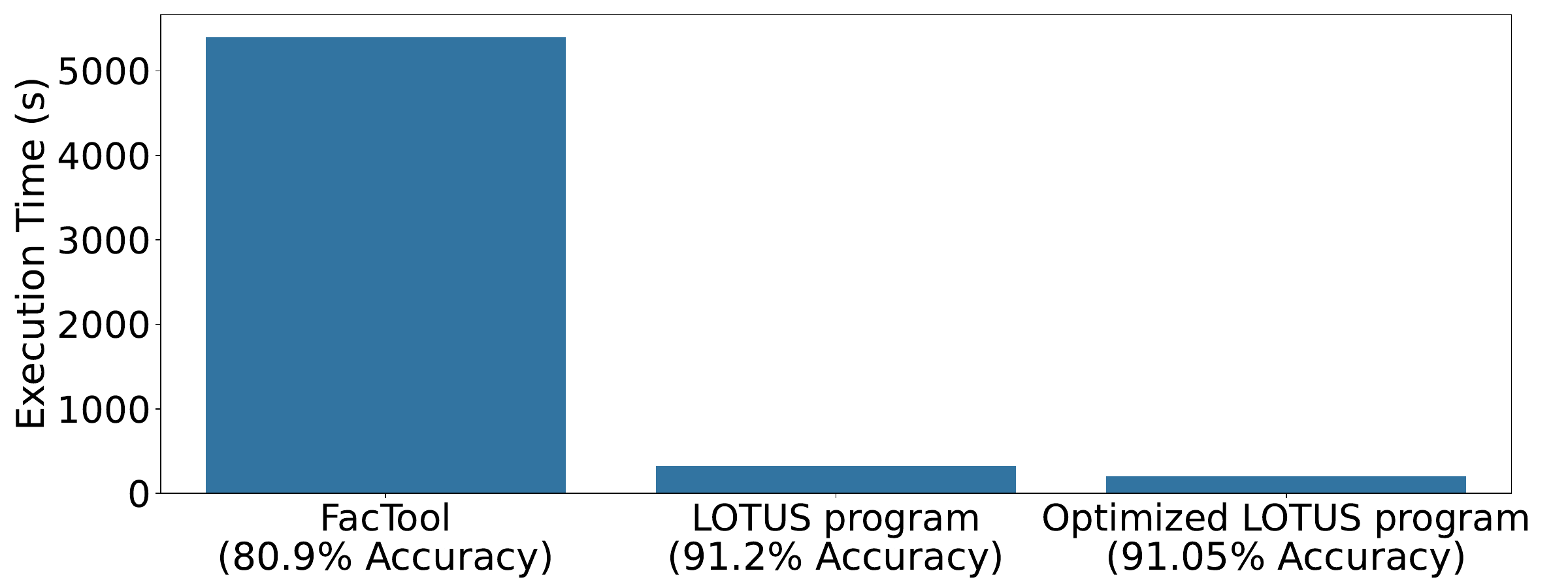}
  \vspace{-.6cm}
  \caption{\small Execution time (y-axis) and accuracy, shown in parentheses, for 3 fact-checking implementations: (i) FacTool~\cite{chern_factool_2023}, a recent state-of-the-art research work, (ii) a short, un-optimized \sys program, and (iii) the same \sys program optimized with accuracy guarantees. Section~\ref{sec:eval} provides our full methodology.}
  \label{fig:factchecking_plot}
   \vspace{-.3cm}
\end{figure}

Unfortunately, existing systems are insufficient for bulk semantic processing, either limiting their expressiveness to row-wise LLM execution models or focusing on empirical optimizations without any formalism or accuracy guarantees, leading to unexpected behaviors and frequent failures. 
First, several systems provide logically row-wise LLM operators, either in the form of a general-purpose AI user-defined functions (UDF)~\cite{motherduck_introducing_nodate, williamdassafmsft_intelligent_2024, noauthor_snow_nodate, vertexai, databricks, noauthor_large_nodate, liu_optimizing_2024} or specialized row-wise operators~\cite{liu_suql_2024, liu_declarative_2024, lin_towards_2024} for specific applications, such as conversational chat, data cleaning and extract-transform-load (ETL) tasks. These programming models fail to support a range of LLM-based transformations \emph{across} rows, such as ranking, grouping or joining records. Alternatively, more recent LLM-based analytics systems, such as DocETL~\cite{shankar_docetl_2024} and UQE~\cite{dai_uqe_2024}, empirically optimize LLM-based operations with no accuracy guarantees, lacking any formalism to define correct behavior, which hinders their robustness and usability, as we show in Section~\ref{sec:eval}. These obstacles highlight a core challenge of integrating semantic-based processing within reliable query systems due to the inherent ambiguity of natural language instructions and LLM outputs.


\begin{figure*}[th]
  \centering
  \includegraphics[width=\linewidth]{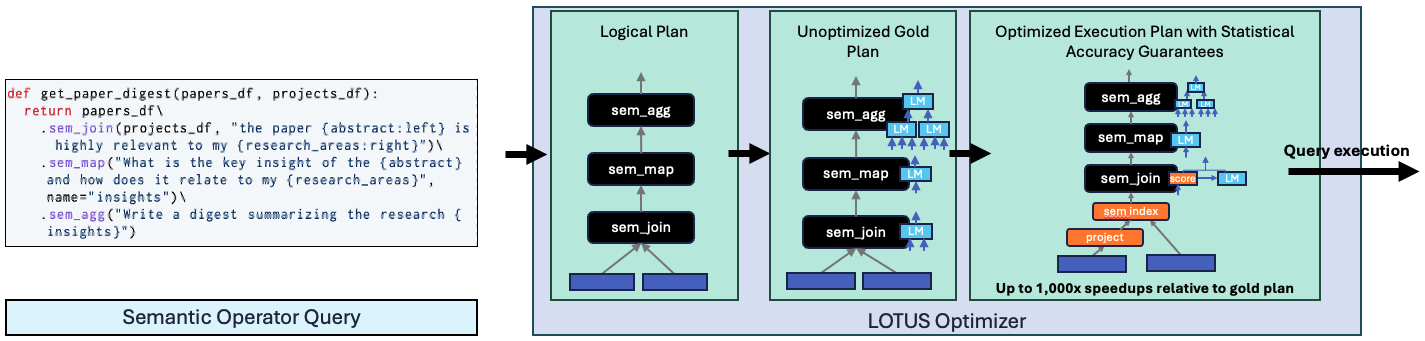}
  \caption{\small An example semantic operator query is shown on the left. The user function takes two two datasets and uses a series of semantic operators to to create a summary of relevant research papers from papers\_df based on the user's research areas from projects\_df. Given a user's query, LOTUS defines a logical plan, which it then converts to an un-optimized gold plan, that specifies how to orchestrate the LM over the data. LOTUS then generates an optimized execution plan with statistical accuracy guarantees relative to the gold plan. These optimized plans empirically demonstrate up to $1,000\times$ speedups. Finally, LOTUS executes the optimized plan and returns the results to the user.}
    \label{fig:example_code}
\end{figure*}

Towards robust, declarative query systems for bulk semantic processing, we propose the \emph{semantic operator model}, which extends the relational model with AI-based operations. Semantic operators provide the first formalism for general-purpose and multi-row, AI-based transformation, including semantic filters, joins, top-k rankings, aggregations, and projections. Each operator takes a concise natural language signature, given by the programmer, and its behavior is fully specified by a tractable, high-quality "gold" algorithm, which indicates how to orchestrate the underlying AI model over the data. 
Our optimization approach then exploits the rich design space of semantic operator execution plans to reduce cost, while providing \emph{statistical accuracy guarantees} with respect to the gold algorithm (i.e., guarantees that the output of the optimized operator will be similar to that of the gold algorithm). 
We propose several novel optimizations for the semantic filter, join, top-k and group-by operators.
Our methods include approximation algorithms, building on statistical techniques used in prior works~\cite{kang_accelerating_2021, kang_approximate_2020} with novel and efficient proxy scorers using small LLMs or semantic embeddings. 
We implement semantic operators in \sys 
(\textbf{L}LMs \textbf{O}ver \textbf{T}ables of \textbf{U}nstructured and \textbf{S}tructured data), an open source system that exposes these operators through a simple DataFrame-based programming interface.

\begin{footnotesize}
    
\begin{table*}[!t]
\newcolumntype{C}[1]{>{\raggedright\arraybackslash}p{#1}}
\renewcommand{\arraystretch}{1.4} 
\centering
\caption{\small Summary of Key Semantic Operators. $T$ denotes a relation, $X$ and $Y$ denote arbitrary tuple types. $l$ denotes a parameterized natural language expression (``langex`` for short). Each operator may permit additional optional parameters, including accuracy targets.}
\vspace{-.3cm}
\begin{tabular}{C{4.1cm}C{5cm}C{4.1cm}C{4.2cm}}
\toprule
\textbf{Operator} & \textbf{Description}  & \textbf{Definition} &  \textbf{Gold Algorithm}\\
\midrule
$\mathit{sem\_filter}(l\textit{: } X \rightarrow \mathit{Bool})$ 
& Returns the tuples that pass the langex predicate. 
& $\{t_i \in T | l_M(t_i) = 1\}$
& Compute $M(t_i, l) \forall t_i\in T$ \\

$\mathit{sem\_join}(t\textit{: } T \comma l\textit{: } (X, Y) \rightarrow \mathit{Bool})$ 
& Joins a table against a second table $t$ by keeping all tuple pairs that pass the langex predicate. 
& $\{(t_i, t_j) | l_M(t_i, t_j) = 1, t_i \in T_1, t_j\in T_2 \}$
& Compute $M(t_i, t_j, l) \forall t_i\in T_1, t_j\in T_2$ \\

$\mathit{sem\_agg}(l\textit{: } T[X] \rightarrow \mathit{X})$ 
& Aggregates input tuples according to the langex reducer function.
& $l_M(t_1, ..., t_n) \forall t_1, .., t_n \in T$
& Perform a reduce algorithm, recursively computing $a_{i+1, j} = M(a_{i, f(j)}, ..., a_{i, f(j)+n'}, l)$, $a_{0, j} = M(t_{f(j)}, ..., t_{f(j)+n'}, l)$ 
\\

$\mathit{sem\_topk}(l\textit{: } T[X] \rightarrow Seq[X]\comma k\textit{: }int)$ & Returns an ordered list of the $k$ best tuples according to the langex ranking criteria.
& $\langle t_1, ..., t_k\rangle \textit{ st } \forall (t_i, t_j), i < j \implies  l_M(t_i, t_j) = \langle t_i, t_j\rangle $
& Perform top-k sorting algorithm using pairwise comparisons, $M(t_i, t_j, l)$\\

$\mathit{sem\_group\_by}(l\textit{: } X \rightarrow Y\comma C\textit{: } int)$ & Groups the tuples into $C$ categories based on the langex grouping criteria. 
& $\underset{\{\mu_1, ...\mu_C\}, \mu_i \in V^\mathbb{N}}\argmax\textit{\space} \underset{t_i\in T}\sum \underset{j\in 1\ldots C}\max l_{M}(t_i, \mu_j)$
& Obtain centers $\mu_1, ...\mu_C$ with a clustering algorithm, and perform pointwise assignments $M(t_i, \mu_1, ..., \mu_C) \forall t_i\in T$\\

$\mathit{sem\_map}(l\textit{: } X \rightarrow Y)$ & Performs the projection specified by the langex. 
& $\{l_M(t_i), \forall t_i \in T\}$
& Compute $M(t_i, l) \forall t_i\in T$\\

\bottomrule
\label{tab:operators}

\end{tabular}
\vspace{-1.0cm}

\end{table*}

\end{footnotesize}

Figure \ref{fig:factchecking_plot} illustrates the power of \sys' declarative programming model and optimized query engine. For a fact-checking task on the FEVER dataset \cite{thorne_fever_2018},  we  write an intuitive \sys program to re-implement a recent state-of-the-art fact-checking pipeline, FacTool~\cite{chern_factool_2023}, which was originally implemented in over 750 lines of code. Our un-optimized \sys program composes $3$ semantic operators (filters, maps and joins)  and achieves  $12.7\%$ higher accuracy on this task. \sys also provides query efficiency by default, yielding $28\times$ lower execution time than the original FacTool implementation.
Furthermore, the optimized \sys program, configured with $0.9$ accuracy targets and a $0.2$  allowed error probability,  runs $1.7\times$ faster and obtains $99.8\%$ accuracy compared to the un-optimized \sys program.

We systematically evaluate \sys on four real bulk-semantic processing applications:  fact-checking, biomedical multi-label classification, search, and topic analysis. 
These include applications expressible and studied by recent LLM-based analytics systems~\cite{shankar_docetl_2024, dai_uqe_2024}, where we compare against those systems, as well as others, where we compare against hand-designed pipelines from the AI literature. 
We demonstrate that the semantic operator model is effective, capturing state-of-the-art AI pipelines in a few operator calls that reliably match or exceed quality of recent LLM-based analytics systems by up to $100\%$, while offering accuracy guarantees and running up to $3.6\times$ faster than the best-quality baselines. 
Furthermore, our gold algorithms provide high-quality results, and our optimizations speed up query execution by up to $1,000\times$ with respect to them.
Specifically, our \sys program re-implements FacTool's recent state-of-the-art fact-checking pipeline~\cite{chern_factool_2023}, as shown in Figure \ref{fig:factchecking_plot}, and achieves $12.5\%$ higher accuracy  with $28\times$  lower execution time on the FEVER dataset ~\cite{thorne_fever_2018}, outperforming alternative LLM-powered analytics systems on both accuracy and efficiency. In the extreme multi-label classification task on the BioDEX dataset~\cite{doosterlinck_biodex_2023}, \sys reproduces state-of-the art result quality~\cite{doosterlinck_-context_2024} and the LLM-powered analytics baselines, while providing  accuracy guarantees and $1,000\times$ speedups with respect to our gold algorithm. In the search and ranking application, \sys allows a simple composition of operators to achieve $3 -  100\%$ higher nDCG@10 than the next best performing baselines, while also providing query efficiency, with $1.67 - 10\times$ lower execution time than alternative high-quality algorithms. In the topic analysis task, \sys efficiently processes hundreds of ArXiv papers to discover a taxonomy of key topics in dozens of seconds, while once again providing configurable speedups with statistical accuracy guarantees.

Overall, our main contributions are the following:
\begin{itemize}
    \item We propose semantic operators, the first formalized programming model for general-purpose AI-based transformations with natural language specifications. 
    \item We define an optimization framework for semantic operators with statistical accuracy guarantees, and novel optimizations for several operators, including semantic filters, joins, top-k and group-by, yielding up to $1,000\times$ speedups.
    \item  We extensively evaluate the expressiveness of the semantic operator model, the quality of our gold algorithms, and the benefit of our optimizations, including comparisons to DocETL, UQE, and AI UDFs. 
\end{itemize}

\section{The Semantic Operator Model}
\label{sec:operators}

Semantic operators provide a declarative interface for semantic-based manipulation and access to data.  Extending the data independence model of relational systems~\cite{codd_relational_1970}, these new \emph{AI-based} operators introduce \emph{model-data independence}, which separates the application logic from the underlying ML-based algorithm and access pattern that specifies how data is passed to model invocations. 
Unlike relational operators, semantic operators are parametrized by natural language expressions and rely on AI-based computation, making their behavior inherently ambiguous and imprecise. We address this key challenge by presenting a formalism for defining semantic operators, their behavior and correct optimizations, in this section.
In Section~\ref{sec:impl}, we build on this formalism to provide optimized execution plans with accuracy guarantees, similar to relational query optimizers.

\subsection{Defining Semantic Operators}
 
\heading{Definition.} \emph{A semantic operator is a declarative transformation over one or more datasets, parameterized by a natural language expression. Each semantic operator can be implemented by many AI-based algorithms, and its behavior is defined with respect to a gold algorithm.}

Table~\ref{tab:operators} lists a core set of semantic operators, which cover common semantic transformations in real-world applications and mirror key transformations in relational operators.
Specific systems may, of course, provide additional semantic operators beyond the ones we discuss here. 
Each semantic operator takes a \emph{parameterized natural language expressions} (\emph{langex} for short), which are natural language expressions that specify a function over one or more attributes. As Figure~\ref{fig:example_code} demonstrates, the langex signature varies for different semantic transformations. The figure shows a \sys program, 
which performs a \verb|sem_join|, \verb|sem_map| and \verb|sem_agg| over datasets of research papers and user projects.
While the \verb|sem_join| langex signature provides a natural language \emph{predicate}, the \verb|sem_map| langex signature indicates a natural language projection, and the \verb|sem_agg| langex is a commutative, associative \emph{aggregator} expression, which here indicates a summarization task over abstracts. 

We provide a high-level definition of each semantic transformation with respect to the user langex and a world model\footnote{In practice, the world model, $M$ may be the strongest LLM a practioner has available.}, $M$, which captures a probability distribution over the vocabulary $V$. For example, semantic filter returns $\{t_i \in T | l_M(t_i) = 1\}$, where $T$ is the input relation and $l_M(t_i)$ represents a natural language predicate evaluated on tuple $t_i$ with model M. 
Notably, the definition of each semantic operator  can be implemented by multiple AI-based algorithms, and different decisions as to how to invoke the model over the relation have consequences on the algorithm's result quality. 
Thus, to fully specify the ground truth behavior of each semantic operator, we define a \emph{gold algorithm}, which is a computable and tractable ML algorithm that produces high-quality results and specifies the model's access pattern over the data, $M(\cdot)$.

\subsection{Defining Correct Optimizations for Semantic Operators}
Semantic operators create a rich design space of diverse execution plans.
While gold algorithms define a tractable, high-quality implementation, they are often expensive, with the complexity of LLM calls scaling linearly or quadratically in the dataset cardinality. For example, the semantic filter's gold algorithm entails invoking the LLM in a linear pass over each tuple. However, an alternate AI-based algorithm may reduce cost, while offering \emph{close results}. In the case of semantic filters, \emph{close results}, can be measured according to accuracy metrics, in this case recall and precision, and guaranteed with high probability compared to the accuracy obtained by the gold algorithm. As such, we define correct optimizations for semantic operators\footnote{Our optimization approach can naturally extend to multi-operator execution plans by composition. In multi-operator programs, the user may configure an accuracy target and failure probability per operator, or for the full program, in which case the optimizer assigns a target accuracy and failure budget to each operator.} below.

\heading{Definition.} \emph{A correct optimization for a given semantic operator, and a gold algorithm for that operator, reduces cost while providing statistical accuracy guarantees with respect to the gold algorithm. Specifically, the optimization should ensure an accuracy target, $\gamma$, is met with probability $1-\delta$.}

In general, optimized semantic operator plans allows for both lossless optimizations and approximations with respect to a gold algorithm. This formulation builds on approaches in approximate query processing and assumes some error is often tolerable, which we find (Section~\ref{sec:eval}) is reasonable for achieving high-quality results, for semantic processing, which is inherently non-exact.

\subsection{Core Semantic Operators}

We now overview several core semantic operators, providing the operator's definition and at least one gold algorithm for each. We discuss design decisions for our gold algorithms based on state-of-the-art algorithms studied in the AI literature, well-known failure cases, such as long-context challenges~\cite{liu_lost_2023}, and our empirical evaluation in Section~\ref{sec:eval}.

\heading{Semantic Filter} is a unary operator over the relation $T$ and returns the relation $\{t_i \in T | l_M(t_i) = 1\}$, where the langex provides a natural language predicate over one or more attributes.

\headingtwo{Gold Algorithm.} Our gold algorithm runs batched LLM calls over all tuples in relation $T$. Each model invocation, $M(t_i, l)$, prompts the LLM with a single tuple $t_i\in T$, the langex predicate, and an operator-specific instruction to generate a boolean value. This simple choice avoids well-studied long-context issues~\cite{liu_lost_2023} by processing rows independently rather than batching multiple tuples within a single prompt invocation.

\heading{Semantic Join} provides a binary operator over relations $T_1$ and $T_2$ to return the relation $\{(t_i, t_j) | l_M(t_i, t_j) = 1, t_i \in T_1, t_j\in T_2 \}$. Here the langex is parameterized by the left and right join keys and describes a natural language predicate over both.

\headingtwo{Gold Algorithm.} 
The gold algorithm implements a \emph{nested-loop join pattern}, performing a single predicate evaluation over a tuple pair with each model invocation $M(t_i, t_j, l) \forall t_i\in T_1, t_j\in T_2$, similar to filters This yields an $O(N_1 \cdot N_2)$ LM call complexity, where $N_1$ and $N_2$ are the table sizes of the left and right join tables respectively.
This quadratic join algorithm is suitable for small join tables but scales poorly for large tables.

\heading{Semanic Top-k} imposes a ranking\footnote{This definition implies that $l_M$ imposes a total and consistent ordering. However, this definition can also be softened to assume partial orderings and noisy comparisons with respect to model $M$.} over the relation $T$ and returns the ordered sequence, $\langle t_1, ..., t_k\rangle  \textit{ st } \forall (t_i, t_j), i < j \implies  l_M(t_i, t_j) = \langle t_i, t_j\rangle$. Here, the langex signature provides a general ranking criteria according to one or more attributes.

\headingtwo{Gold Algorithm.}
Two important algorithmic design decisions for the semantic top-k include how to implement LLM-based comparison and how to aggregate ranking information from these comparisons. Our gold algorithm uses pairwise LLM comparisons for the former, and a quick-select top-k algorithm~\cite{hoare_algorithm_1961} for the latter. We briefly describe the reason for these choices and alternatives considered. 
Our decisions build on prior works, which have studied LLM-based passage re-ranking~\cite{desai_calibration_2020, drozdov_parade_2023, liang_holistic_2022, ma_zero-shot_2023, pradeep_rankvicuna_2023, pradeep_rankzephyr_2023, qin_large_2024, sachan_improving_2022, sun_is_2023} and ranking with noisy comparisons~\cite{shah_simple_2016, braverman_noisy_nodate} with the goal of achieving high quality results in a modest complexity of LLM calls or comparisons.

First, pairwise-prompting methods offer a simple and high-quality approach that feeds a single pair of tuples to each LLM invocation, $M(t_i, t_j, l)$, prompting the model to compare the two inputs and output a binary label. The two main classes of alternatives are point-wise ranking methods~\cite{desai_calibration_2020, drozdov_parade_2023, liang_holistic_2022, sachan_improving_2022, wu_stark_2024}, and list-wise ranking methods~\cite{ma_zero-shot_2023, pradeep_rankvicuna_2023, pradeep_rankzephyr_2023, sun_is_2023}, both of which have been shown to face quality issues~\cite{sun_is_2023, qin_large_2024, desai_calibration_2020}. We verify these limitations in Section~\ref{sec:eval}. In contrast, pairwise comparisons have been shown to be effective and relatively robust to input ordering~\cite{qin_large_2024}. 

In addition, we consider several possible rank-aggregation algorithms, including quadratic sorting algorithms, a heap-based top-k algorithm and a quick-select-based top-k ranking algorithm. All three alternatives represent candidate gold algorithms. Our evaluation in Section~\ref{sec:eval} demonstrates that each of these sorting algorithms yield high-quality results, comparable to one another. However, the quick-select-based algorithm offers an efficient implementation with at least an order of magnitude fewer LLM calls than the quadratic sorting algorithm and more opportunities for efficient batched inference, leading to lower execution time, compared to a heap-based implementation. The quick-select top-k algorithm proceeds in successive rounds, each time choosing a pivot, and comparing all other remaining tuples to the pivot tuple to determine the rank of the pivot. Because each round is fully parallelizable, we can efficiently batch these LLM-based comparisons before recursing in the next round.

\heading{Semantic Aggregation} performs a many-to-one reduce over the input relation, returning $l_M(t_1, ..., t_n),  \forall t_1, .., t_n \in T$. Here, the the langex signature provides a commutative and associative aggregation function\footnote{We note that the ordering of inputs within an LLM prompt invocation can in fact affect results quality for some tasks. Thus, programmers may wish to override the commutativity and associativity assumptions of the semantic aggregation langex. For this, the LOTUS API exposes a partioner function which groups and orders inputs within LLM prompts according to an arbitrary user-specification.}, which can be applied over any subset of rows to produce an intermediate results. We note that the langex iteself is model-agnostic, assuming infinite context. Managing finite context limits of the underlying model $M$ is an implementation detail left to the system.

\headingtwo{Gold Algorithm.}
Our gold algorithm must efficiently orchestrate the LLM to manage long context inputs, which may degrade result quality~\cite{liu_lost_2023} or overflow the underlying model's context length. We select a hierarchical reduce pattern for the gold algorithm. Our choice builds on the LLM-based summarization pattern studied by prior research works~\cite{wu_recursively_2021, chang_booookscore_2024, adams_sparse_2023} 
and deployed systems~\cite{llamaindex, langchain}, which we briefly overview. 

Prior works primarily implement one of two aggregation patterns: either a fold pattern, which performs a sequential, linear pass over the data, while iteratively updating an accumulated partial answer, or a hierarchical reduce pattern, which recursively aggregates the input data to produce partial answers until a single answer remains. 
Both represent candidate gold algorithms, however, the hierarchical pattern has been shown to produce higher quality results for commutative, associative aggregation tasks, like summarization, in prior work~\cite{chang_booookscore_2024} and allows for greater parallelism during query processing, making it our default choice.

\heading{Semantic Group-by} takes a langex that specifies a projection from a tuple to an unknown group label, as well as a target number of groups, which specifies the desired granularity of group labels. As a example, a user might group-by the topics presented in a set of ArXiv papers, wishing to find 10 key groups. The group-by operator must \emph{discover} representative group labels and assign a label to each each tuple. In general, performing the unsupervised group discovery is a clustering task, which is NP-hard~\cite{dasgupta_hardness_nodate}. Clustering algorithms over points in a metric space typically optimize the potential function tractably using coordinate descent algorithms, such as k-means. For the semantic group-by, the clustering task is over unstructured fields with a natural language similarity function specified by $l_M(t_i, \mu_j)$, which imposes a real-valued score between a tuple $t_i$ and a candidate label $\mu_j$. This operator poses the following optimization problem: 

$$\argmax_{\{\mu_1, ...\mu_C\},\textit{\space}\mu_i \in V^\mathbb{N}} \sum_{t_i\in T} \max_{j\in 1\ldots C} l_{M}(t_i, \mu_j)$$

\noindent where $\mu_1, ..., \mu_C$ are $C$ group labels, each consisting of tokens in the vocabulary, $V$.

\headingtwo{Gold Algorithm.}
Since this operator, by definition, entails a clustering problem, known to be NP-hard~\cite{dasgupta_hardness_nodate}, our gold algorithm instead offers a computable implementation to obtain group labels using a tractable clustering algorithm to discover group labels followed by a point-wise classification step. Both steps use a linear LLM pass over the relation. 

Specifically, the first stage performs an LLM-based clustering algorithm to discover centers $\mu_1, ..., \mu_C$, given user langex. Our gold algorithm first performs a semantic projection, prompting the LLM to predict a label for each input tuple. Then, we embed these candidate labels and perform an efficient vector clustering using a k-means algorithm to construct $C$ groups. For each group, we top-k sample tuples based on centroid-similarity scores, and use a semantic aggregation that prompts the model to generate an appropriate label over each group. In the second stage of our gold algorithm, we use the $C$ generated labels, $\mu_1, ...\mu_C$ and invoke the model to perform point-wise assignments $M(t_i, \mu_1, ...\mu_C), \forall t_i\in T$.

\section{Optimized Execution Plans for Semantic Operators} 
\label{sec:impl}

In this section, we present novel optimizations for several costly operators, including semantic filters, joins, top-k and group-by. Each optimization provides statistical accuracy guarantees with respect to the gold algorithm, building from Section~\ref{sec:operators}.
While this paper focuses on novel optimizations for several expensive semantic operators, we envision a rich space of future work exploring new optimizations and applications of traditional optimizations to semantic operator programs. For example, several prior works demonstrate performance gains in both logical query plan optimizations (e.g. operator re-ordering~\cite{liu_suql_2024, liu_declarative_2024, lin_towards_2024, lu_accelerating_2018}) and other general LM-approximation techniques (e.g. code synthesis~\cite{arora_language_2023, liu_declarative_2024} and prompt adaptation~\cite{chen_frugalgpt_2023}).

\subsection{Optimizing Semantic Filter}

\begin{footnotesize}

\RestyleAlgo{ruled} 

\SetKwComment{Comment}{// }{}

\begin{algorithm}[t!] 
\caption{SEM-FILTER($T$, $l$, $M(x)$, $A(x)$, $\gamma_R$, $\gamma_P$, $\delta$)}
\label{alg:approx_sem_filter}

\KwIn{Relation $T$, langex predicate $l$, oracle model $M(x)$, proxy model $A(x)$, recall target $\gamma_R$, precision target $\gamma_P$, error probability $\delta$}
\KwOut{Filtered relation $T'$}

$s \gets \textit{sample\_size}$\\ 
$S \gets \texttt{ImportanceSample}(T, A(x), s)$\\
$M_S \gets \{M(t) : t \in S\}$\\
$A_D \gets \{A(t) : t \in D\}$\\
$A_S \gets \{A(t) : t \in S\}$\\
$\tau_+ \gets \texttt{PT\_threshold\_estimation}(S, l, M_S, A_S, \gamma_P, \delta / 2)$\\
$\tau_- \gets \texttt{RT\_threshold\_estimation}(S, l, M_S, A_S, \gamma_R, \delta / 2)$\\
$\tau_+ \gets \max(\tau_+, \tau_-)$\\
$T' \gets \emptyset$\\

\Comment{Evaluate predicate for each tuple}
\For{$t \in T$}{
    \uIf{$A(t) \geq \tau_+$}{
        $T' \gets T' \cup \{t\}$
    }
    \uElseIf{$A(t) \geq \tau_-$}{
        \uIf{$M(t)$}{
            $T' \gets T' \cup \{t\}$
        }
    }
}
\Return{$T'$}

\end{algorithm}
\end{footnotesize}

\label{subsubsec:filter_cascade} We provide an approximation for semantic filters that obtains recall and precision targets $\gamma_R$ and $\gamma_P$ with respect to the gold implementation with probability $1-\delta$. To achieve reduced cost we leverage the cheaper, but less accurate proxy model $A(t, l)$, which outputs a score indicating whether the tuple $t$ passes the predicate.
This idea is inspired by prior works~\cite{viola_rapid_2001, yue_large_2024, chen_frugalgpt_2023, kang_noscope_2017, kang_approximate_2022, yue_large_2024}, which leverage model cascades for different problem settings. Specifically, several works study cascades in the video analytics setting with vision models, which have significantly different properties than LLMs, requiring different proxy scoring mechanisms. Other works study cascades for LLMs, but use heavy-weight scoring mechanisms to decide whether to use the proxy model and do not provide accuracy guarantees. In contrast, we focus on providing accuracy guarantees for the constrained problem of applying cascades for filters, which allows us to use lighter-weight scoring functions than the more general case studied in prior works.

In general, we cannot assume the proxy model is accurate. In fact, the proxy model may perform very poorly for some tasks and records. The goal of our algorithm is to automatically discover the quality of the proxy \emph{at query time} by sampling and comparing to the gold algorithm with the oracle model. Our algorithm will exploit the proxy when it is likely to provide high-quality outputs, and otherwise defer the oracle model. To do this we must learn a decision rule, with learned thresholds on the proxy scores using sampling. As prior work~\cite{kang_approximate_2020} discusses, such sampling-based algorithms introduce multiple-hypothesis testing problems, requiring statistical corrections using confidence intervals, which we carefully apply in our algorithm.

We consider a small LLM as the proxy model and generate scores $A(t)$ using log-probabilities\footnote{This optimizations assumes access to log-probabilities, which is available in common model providers, such as OpenAI, and serving systems, such as vLLM. In the absence of available log-probabilities, alternative uncertainty quantification methods, such as prompt-based methods, may be more suitable.} corresponding to the \verb|True| or \verb|False| output tokens of the proxy model's predicate evaluations, re-scaling by the quantiles over all generated log-probabilities over the relation. 
Algorithm~\ref{alg:approx_sem_filter} shows the full procedure for performing the approximate semantic filter. We begin by collecting a sample of tuples $S$, and labeling each sample with the oracle and proxy models. The sampling procedure uses importance sampling and defensively mixes a uniform sample following prior work~\cite{kang_approximate_2022}. Using the central limit theorem and the normal approximation on the distribution of sample statistics, we then learn a decision rule by finding thresholds $\tau_+$ and $\tau_-$, which will respectively ensure the precision target and recall target are met, each with an error probability of $\delta/2$. The statistical sub-procedures to choose each threshold follows prior work~\cite{kang_approximate_2022}, but applies them to this new setting where we must ensure \emph{both} targets are met, requiring a correction for hypothesis testing and multiple failure modes. 
Finally, using the learned decision rule, our algorithm proceeds to process each tuple in the relation. If the tuple's proxy score is at least $\tau_+$, we mark it as passing the predicate. If the tuple's proxy score is less than or equal to $\tau_-$, we mark it as failing the predicate, and for tuples with proxy scores between $\tau_+$ and $\tau_-$, we resort to the oracle model to provide a label.

\subsection{Optimizing Semantic Join}

Similar to semantic filters, we provide an approximation for semantic joins that obtains recall and precision targets $\gamma_R$ and $\gamma_P$ with respect to the gold algorithm with probability $1-\delta$. Due to the expensive quadratic scaling of the nested-loop join, rather than using a small LLM as the proxy model, we consider an even cheaper proxy based on semantic similarity scores using embeddings. 
We leverage two possible proxy algorithms and dynamically choose the lowest cost one. 

The first approximation, \textbf{sim-filter} produces a proxy score $A_1(t_i, t_j)$, $t_i\in T_1, t_j\in T_2$ based on embedding similarity. We perform batched similarity search between the right and left join keys and re-calibrate similarity scores according to their quantiles to obtain proxy scores. 
This approximation is likely to perform efficiently when tuple pairs with high semantic similarity between the the right and left join key are more likely to pass the predicate. This correlation phenomenon between predicate matches and embedding similarity has been studied by prior works~\cite{patel_acorn_2024} and is not always present. Thus, we introduce an alternative approximation, often suitable when correlation is not present.

The second approximation, \textbf{project-sim-filter} first performs a projection over the left join key and then uses the projected column to compute proxy scores $A_2(t_i', t_j)$ uses the embedding similarity between the projected value $t_i'$ and $t_j$. As before, we also re-calibrate the proxy scores taking their quantiles.
Intuitively, performing the projection is useful when tuple pairs that pass the user's natural language predicate exhibit low semantic similarity. The projection step invokes the LLM over each tuple in left join table, prompting it to predict values for the right join key. Notably, this LLM projection is ungrounded, i.e., it is done without knowledge of the right join key's attribute domain and can thus be performed in a fully parallelized semantic map operation. 
Figure~\ref{fig:example_llmjoin2} provides an example of a semantic join between a table of papers and datasets, where the predicate evaluates whether a given paper abstract uses a specific dataset. Here, the map step would invoke the LLM over each abstract, instructing the model to output the dataset used, conditioned only on the abstract. 

To dynamically choose between these two approximations, we follow a procedure similar to Algorithm~\ref{alg:approx_sem_filter} used for semantic filters. We begin by importance sampling to collect the sample $S$ and construct the set of oracle labels, $O_S$, over the sample. We then obtain tresholds $\tau_+$ and $\tau_-$ independently for each approximation algorithm using their respective proxy scores, $A_1$ and $A_2$. We use the learned thresholds for each candidate plan to then determine the exact oracle cost needed to execute either algorithm, and we take the least cost plan with its associated learned thresholds to proceed to evaluate the predicate for each tuple pair.

\begin{figure}[t]
    \centering
    \vspace{-.2cm}
    \begin{lstlisting}[language=Python]
papers_df.sem_join(dataset_df, "The paper {abstract:left} uses the {dataset:right}.", recall_target=0.9, precision_target=0.9, delta=0.1)
\end{lstlisting}
    \vspace{-.5cm}
    \caption{\small Example sem\_join for matching papers and datasets.}
    \label{fig:example_llmjoin2}
    \vspace{-.5cm}
\end{figure}

\subsection{Optimizing Semantic Group-by}
We provide an efficient approximation that guarantees a classification accuracy target, $\gamma$, is met with probability $1-\delta$. We follow the clustering algorithm in the first stage of the gold algorithm to discover centers $\mu_1, ...\mu_C$. We then use a proxy-based approximation for point wise assignments in the second stage of the algorithm. Similar, to semantic joins, we consider semantic similarity scores as a cheap proxy, although other proxy models are also feasible. Specifically, we leverage the embeddings constructed during the clustering stage and compute similarity scores between the the candidate label of each tuple and the discovered centers  $\mu_1, ...\mu_C$. The proxy score of the tuple $t_i$ with a predicted label, $t_i'$, for a discovered center $\mu_j$ is given by $A(t_i, \mu_j) = sim(t'_i, \mu_j)$. Our goal is then to learn a threshold, $\tau$, such that if the proxy score between a tuple and center is greater than $\tau$, we return the center as label for the tuple, and otherwise we resort to the more expensive LLM-based classification procedure. We accomplish this by uniform sampling and running the sub-procedure equivalent to the $\verb|PT_threshold_estimation|$ used in Algorithm~\ref{alg:approx_sem_filter} for semantic filters, where the metric we evaluate is classification accuracy over the $C$ groups.

\subsection{Optimizing Semantic Top-k}
We leverage the embedding similarity scores to optimize pivot selection for some queries, while incurring no accuracy loss. This optimization is useful when there exists correlation between the ranking imposed by the user's arbitrary sorting criteria and the ranking imposed by semantic similarity scores. In this case, we can sort tuples based on embedding distances to the user's query, and select the $(k + \epsilon)$-th item, rather than a random item, as the first pivot. This can reduce the number of LLM comparisons required by subsequent rounds in the quick-select algorithm, leading to higher query efficiency at no accuracy loss. In the case of no correlation between the langex-based ranking and similarity-based ranking, this method amounts to random pivot selection, and in the worst case of an adversarial pivot, the algorithm will incur one extra pivot round, which can impact execution time, but not degrade quality.

\section{The \sys Systems}
\label{subsec:llm_transformations}
In this section we describe the \sys system, which implements the semantic operator model as an extension of Pandas~\cite{pandas}. We chose a Pandas-like API in our initial system implementation to make it easy for users to integrate \sys with popular AI libraries in Python. However, semantic operators could also be added to a variety of other data processing APIs and query languages, such as SQL. 
\sys leverages vLLM~\cite{kwon_efficient_2023} to perform efficient batched inference, and uses FAISS~\cite{douze_faiss_2024, johnson_billion-scale_2017} to support efficient vector search for \sys' semantic search, similarity join and cluster operations, with indices stored and maintained locally on disk, by default.

\subsection{Datatypes} 

\sys' data model consists of tables with structured and unstructured fields (e.g., text or images). \sys' semantic operators can take both of these data-types as parameters to a langex.
Additionally, \sys supports \emph{semantic indices} over natural-language text columns to provide optimized query processing. These indices leverage semantic embeddings over each document in the column and capture semantic similarity using embedding distance metrics. Semantic indices can be created off-line with \verb|sem_index| and a specified retriever model, and then loaded from disk using \verb|load_sem_index|.

\subsection{Semantic Operators in \sys}
We now overview the semantic operators supported in the \sys API, which includes the core set of operators described in Section~\ref{sec:operators}, as well as several additional variants provided for convenience.
Each operator takes optional parameters to specify a target accuracy and error probability, which the optimizer will use to transparently perform optimizations.

\heading{Sem\_filter, Sem\_join, \& Sem\_sim\_join.} The \sys API supports \verb|sem_filter| and \verb|sem_join|, both of which take a langex predicate, as described in Section~\ref{sec:operators}. In addition, \sys provides a join variant, \verb|sem_sim_join|, where tuples are matched according to their \emph{semantic similarity}, rather than an arbitrary natural-language predicate. Akin to an equi-join in standard relational algebra, the semantic similarity join is a specialized semantic join, which indicates additional optimization opportunities to the query engine by leveraging vector similarity search. Figure~\ref{fig:example_llmjoin} provides an example of the \verb|sem_join| compared to the \verb|sem_sim_join|, where the user specifies the left and right table join keys, and a parameter $K$.  
The operator performs a left join such that for each row in the left table, the output table will contain $K$ matching rows from the right table with the highest semantically similarity scores. 
\begin{figure}[h]
    \vspace{-.2cm}
    \centering
    \begin{lstlisting}[language=Python]
papers_df.sem_join(papers_df, "The paper {abstract:left} contradicts the claims in {abstract:right}.")
papers_df.sem_sim_join(papers_df, left_on="abstract", right_on="abstract",  K=10)
\end{lstlisting}
    \vspace{-.5cm}
    \caption{\small Example usage of sem\_join and sem\_sim\_join.}
    \label{fig:example_llmjoin}
    \vspace{-.5cm}
\end{figure}

\heading{Sem\_topk \& Sem\_search}. \sys supports a semantic top-k, which takes the langex ranking criteria, as described in Section~\ref{sec:operators}. Programmers can optionally specify a group-by parameter to indicate a subset of columns to group over during ranking, as shown in Figure~\ref{fig:example_semtopk_groupby}. The groupings are defined using standard equality matches over the group-by columns. Additionally, as the figure shows, \sys also provides a top-k variant, \verb|sem_search|, which assumes a semantic similarity-based ranking criteria relative to a natural language query. \sys also exposes advanced relevance-based re-ranking functionality for search, allowing users to specify the \verb|n_rerank| parameter during the semantic search. The semantic search in this case will first find the top-$K$ most relevant documents and then re-rank the top-$K$ found documents to return the top \verb|n_rerank|.

\begin{figure}[h]
    \vspace{-.2cm}
    \centering
    \begin{lstlisting}[language=Python]
papers_df.sem_topk("the {abstract} makes the most outrageous claim", K=10, group_by=[arxiv_domain])
papers_df.sem_search(col="abstract", query="vector databases", K=10)
\end{lstlisting}
    \vspace{-.5cm}
    \caption{\small Example usage of sem\_topk and sem\_search.}
    \label{fig:example_semtopk_groupby}
    \vspace{-.5cm}
\end{figure}

\heading{Sem\_agg} performs an aggregation over the input relation, with a langex signature that provides a commutative and associative aggregation function, as shown by Figure~\ref{fig:example_code}. Similar to \verb|sem_topk|, \sys allows users to specify a group-by parameter.

\heading{Sem\_group\_by} creates groups over the input dataframe according to the langex grouping criteria and target number of groups. By default, the operator discovers suitable group labels, but the user can also optionally specify target labels. This operator is useful both for unsupervised discovery of semantic groups, and for semantic classification tasks.

\heading{Sem\_map \& Sem\_extract} both perform a natural language projection over an existing column. While \verb|sem_map| projects to an arbitrary text attribute, \verb|sem_extract| projects each tuple to a list of sub-strings from the source text. This is useful for applications, such as entity extraction, where finding snippets or verified quotes may be preferable to synthesized answers.

\section{Evaluation}
\label{sec:eval}
We systematically evaluate \sys on four bulk-semantic processing applications from the AI and data literature:  fact-checking, biomedical multi-label classification, search, and topic analysis. 
Some of these applications can be expressed by recent LLM-based analytics systems, like AI UDFs, UQE and DocETL, allowing us to compare against them; while other applications are not supported by the baseline systems, but LOTUS supports them. For each application, we additionally compare against strong hand-designed pipelines from the literature. We evaluate the LOTUS optimizer, comparing the performance of optimizations for semantic filters, joins, top-k and group-by with our gold algorithms. Lastly, we analyze the statistical accuracy guarantees provided by LOTUS' optimizations. Overall, we find the following:
\begin{itemize}
    \item Compared to AI-based analytics systems, like AI UDFs, UQE, and DocETL, LOTUS attains similar or up to $170\%$ higher accuracy, while running up to $3.6\times$ faster and offering statistical accuracy guarantees that the other systems do not provide.
    \item Compared to hand-written pipelines from the AI literature for fact-checking, biomedical classification, and search, LOTUS matches state-of-the-art quality, or exceeds it by up to $100\%$, in programs involving a few semantic operators.
    \item The LOTUS optimizer can substantially reduce cost of semantic operator programs, by up to $1,000\times$ compared to high-quality gold algorithms, and provides accuracy guarantees, which hold across repeated trials.
\end{itemize}

To report controlled latency numbers, we run each baseline and experiment locally on 4 80GB A100 GPUs using Llama-3-70B~\cite{noauthor_introducing_nodate} and E5 embeddings~\cite{wang_text_2024}, with a batch size of 64 running on vLLM~\cite{kwon_efficient_2023}, unless otherwise stated. To run DocETL baselines and reproduce some of their benchmarks~\cite{shankar_docetl_2024}, we additionally use OpenAI~\cite{noauthor_openai_nodate} models, including GPT-4o-mini-2024-07-18, GPT-4o-2024-08-06 model, and text-embedding-3-small, where noted. For our experiments that use OpenAI models , we control for cost by limiting execution to 64-way thread parallelism. For reproducibility, we set temperature to $t=0$ for all methods and baselines, unless otherwise stated.

\subsubsection{Benchmarked Methods}
We briefly overview the main methods we benchmark and tested parameters.

\heading{AI UDFs.} Many database vendors~\cite{vertexai,  databricks, noauthor_large_nodate, motherduck_introducing_nodate, williamdassafmsft_intelligent_2024, noauthor_snow_nodate} now support AI UDFs, which include map-like, row-wise LLM operations and primitives for vector search. To control for serving infrastructure, we implement these programs using LOTUS running on vLLM, and restrict our API to \verb|sem_map|, \verb|sem_search| and \verb|sem_sim_join|.

\heading{UQE.} UQE~\cite{dai_uqe_2024} proposes an optimized LLM-powered filter method, which is relevant to this work. UQE also optimizes non-LLM aggregates (e.g. COUNT) integrated with LLM-based queries, but this is beyond the scope of our evaluation. Since the code is not open-source, we implement the UQE filter in about 100 line of code in python. The UQE filter method takes an LLM call budget and provides a best-effort embedding-based approximation. In contrast, LOTUS provides a general proxy-oracle approximation method, where the proxy may be embedding-based or LLM-based. To provide a fair comparison between UQE and LOTUS, we normalize the \emph{latency budget} when comparing against LOTUS programs that use LLM-based proxies, and we normalize \emph{the LLM call budget} when comparing against LOTUS programs that use embedding-based proxies. We also hyper-parameter tune the mini-batch size used for UQE's active learning algorithm. 

\heading{DocETL.} DocETL~\cite{shankar_docetl_2024} uses an LLM agent as the query optimizer to perform rewrites to programs, allowing programs to specify separate LLMs for the optimizer and execution. Our experiments find that the optimizer often fails, and among runs where the optimizer succeeds, performance varies. Based on personal communication with the authors, we run the DocETL baselines multiple times. We obtain three successful runs and report the average performance of these three successes.
We also try using both Llama-70B as the DocETL optimizer, following our setup for each other baseline, and GPT-4o-mini as the DocETL optimizer, following the DocETL preprint~\cite{shankar_docetl_2024}. However, we find both fail to produce a successful run in Section~\ref{subsec:biodex}.
After reaching out to the authors, we confirm a mistake in the pre-print and find that GPT-4o is required for the DocETL optimizer for our evaluation on BioDEX in Section~\ref{subsec:biodex}.
We also note we found several bugs in the DocETL codebase, which we have done our best to fix and have also coordinated with the authors about. Since the codebase is an evolving artifact, we report the commit number we used for our evaluation\footnote{\url{https://github.com/ucbepic/docetl/commit/93050998077a9eb6fc1ee99dc96d1e0a222a987f}}.

\heading{LOTUS.} We set default accuracy targets of $\gamma=0.9$ and failure probabilities of $\delta=0.2$. We provide a detailed ablation, varying these parameters in Section~\ref{subsec:ablation}. Additionally, our sampling size uses $0.01\%$ of the data, similar to other works~\cite{shankar_docetl_2024}, and a minimum sample size of $s=100$ for smaller datasets, following prior work~\cite{kang_approximate_2020}.

\subsection{Fact-Checking}
\label{subsec:factchecking}
Fact-checking systems determine whether a given claim is correct, typically based on verifiability with a knowledge corpus. 
FacTool~\cite{chern_factool_2023} is a recent open-source research work that provides a  multi-step fact-checking pipeline, involving claim extraction, query generation, tool querying, evidence collection, and verification. We consider the task of using constructing the FacTool pipeline, which was originally written in over 750 lines of code. We use the FEVER \cite{thorne_fever_2018} dataset, a claim verification dataset, for our evaluation. Claim verification is based on a corpus of $~5.5$ million Wikipedia articles, and each claim is labeled with one of three gold labels, "Supported", "Refuted", or "NotEnoughInfo". We merge the latter two labels into a single class, "Not Supported", following prior work~\cite{chern_factool_2023} for our evaluation. To control for our API spending budget, we sample 1,000 claims from the development dataset. 

\heading{Baselines.} 
For all baselines we use ColBERT~\cite{khattab_colbert_2020} as the retreiver model\footnote{FactTool's pipeline, by default, performs retrieval with a Google Search API \cite{noauthor_custom_nodate}. We evaluate the pipeline with both the default retrieval API, and with the open-source ColBERT \cite{khattab_colbert_2020} index. We find that the results are similar, and we report the results using ColBERT for retrieval to hold the retriever model constant with the other baselines.} and Llama-70B served with vLLM. We run FactTool's open source codebase~\cite{noauthor_gair-nlpfactool_nodate} to measure its performance.
The AI UDF baseline uses a simple semantic map-search-map dataflow, following FacTool's pipeline. First each claim is mapped to two search queries. Then the generated queries are used for search over the Wikipedia corpus. Lastly, each claim appended with retrieved context is mapped to truth label, along with a reasoning. We use the same prompts found in FacTool \cite{noauthor_gair-nlpfactool_nodate}, which include 3 demonstrations for generating search queries in the first semantic mapping, and chain-of-thought prompting in the second semantic mapping. The UQE baseline follows the first two steps of the AI UDF baseline, but replaces the final step with UQE's embedding-based LLM-powered filter. Our LOTUS program similarly follows a map-search-filter pipeline and uses a Llama-8B proxy model. To provide a fair comparison, we tune the UQE LLM call budget to normalize execution time with the optimized LOTUS program. Since DocETL does not provide search primitives, we are unable to benchmark it in this application.

\begin{footnotesize}
    
\begin{table}[!t]

\centering
\caption{\small Fact-checking Performance on the FEVER Dataset.}
\vspace{-.3cm}
\begin{threeparttable}
\begin{tabular}{p{2.2cm}p{1.cm}p{1.3cm}p{1.6cm}p{.6cm}}
\toprule
Method & Accuracy & ET (s), with batching & ET (s), no batching & LoC\\
\midrule
FacTool & 80.9 & N/A & 5396.11 & > 750\\
AI UDF: map, search, map & 89.9 & 688.9 & 4,454.2  & < 50\\
AI UDF map, search + UQE filter & 66.0 & 184.4 & 738.3 &  150\\
\sys map, search, filter (unopt.) & 91.2 & 329.1 & 989.0 & < 50\\
\sys map, search, filter (opt.)  & 91.0 & 190.0 & 776.37 & < 50\\
\bottomrule
\vspace{-.3cm}
\end{tabular}
\end{threeparttable}
\label{tab:factchecking}
\vspace{-.3cm}
\end{table}
\end{footnotesize}

\heading{Results.} We report the accuracy of each method, an estimate of lines of code (LoC), and average execution time over 10 runs in seconds, both with and without batching to provide a fair comparison with FacTool, which provides a sequential implementation only. 
Overall, Table~\ref{tab:factchecking} demonstrates that the optimized LOTUS program reproduces state-of-the art accuracy, comparable to that of FacTool, with $7\times$ faster unbatched execution and $28\times$ faster batched execution, and in less than 50 lines of user code, while also outperforming each baseline. 

Comparing the optimized LOTUS program to the AI UDF baseline, we see that both acheive comparable accuracy in relatively few lines of code. However, the LOTUS \verb|sem_filter| can be more heavily optimized compared to the generic AI UDF map operation. This allows the LOTUS program to attain $3.6\times$ faster execution, attributable to the \verb|sem_filter|'s short LLM generations, which leads to cheaper decoding, and our proxy-based optimization for filters. We also observe that the LOTUS program outperforms the UQE program by $38\%$ accuracy at an equivalent latency budget. While our filter optimization adaptively learns when to leverage the proxy in order to meet an accuracy target, UQE's embedding-based approximation provides no accuracy guarantees. 

\begin{figure}[t]
  \centering
  \includegraphics[width=.45\linewidth]{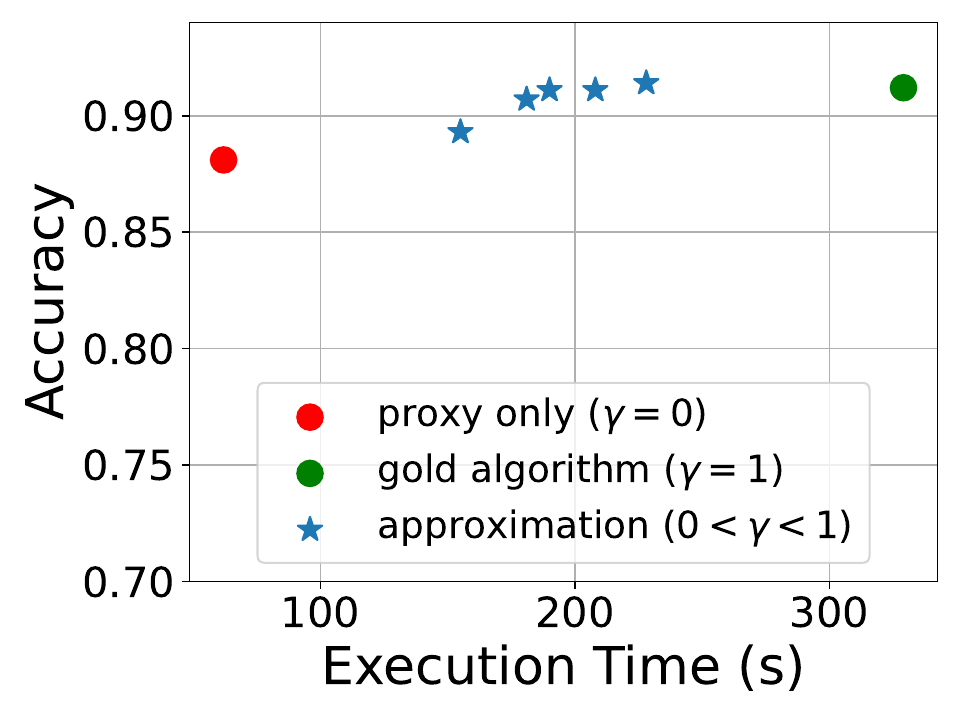}
  \vspace{-.5cm}
  \caption{\small Accuracy vs. execution time (s) for the \sys fact-checking program on the FEVER dataset. We compare the performance using the proxy only (red circle) or oracle only (green circle) for filters to the pipeline implemented using our filter approximation with both models for varied the precision and recall target, at $\delta=0.2$ (stars).}
  \label{fig:factchecking_cascades}
   \vspace{-.3cm}
\end{figure}

Finally, we compare the optimized LOTUS program to the un-optimized LOTUS program. We observe that the optimized \sys program achieves $99.8\%$ accuracy relative to the un-optimized one, while reducing batched execution time by $1.7\times$. Figure~\ref{fig:factchecking_cascades} highlights the diverse performance trade-offs achievable by our \verb|sem_filter| approximation. The plot shows the accuracy and execution time of the LOTUS program using the gold algorithm with Llama 70B, or using the proxy only, shown respectively by the green and red circles.
By varying the filter recall and precision target, our cascade-based approximation offers diverse trade-offs between accuracy and execution time, shown by the stars.

\subsection{Biomedical Multi-label Classification}
\label{subsec:biodex}

Biomedical classification entails processing complex, lengthy patient documents. We consider the extreme multi-label classification task on the BioDEX Dataset~\cite{doosterlinck_biodex_2023}, which consists of a corpus of $65,000$ biomedical articles, and ground-truth expert-created drug safety reports constructed from each article. The task is to classify the patient's drug reactions, given each medical article.  Notably, there are $~24,000$ possible drug-reaction labels, making this task an extreme multi-label classification task. Due to the large number of possible labels, leveraging an LLM to perform inference is difficult, and this setting has been studied in prior works~\cite{doosterlinck_-context_2024}. We show below that this task can be efficiently modeled and optimized using the semantic join. Similar to prior work~\cite{doosterlinck_biodex_2023} we sample 250 patient articles for our evaluation to control for our API spending budget.

\heading{Baselines.} 
The search baseline performs batched similarity search over each patient article and the set of reaction labels. For the baseline LLM-based analytics systems, we implement a LLM-powered join over the patient articles and set of candidate reaction labels, followed by a ranking step, if supported. The AI UDF program entails a naive, nested-loop join pattern using row-wise LLM operators; however we note this is prohibitively costly and we report estimated latency. While UQE does not explicitly study joins, we implement a UQE join using its optimized filter. The DocETL program uses a join and reduce, which uses listwise ranking, following the preprint~\cite{shankar_docetl_2024}. We note that we are unable to successfully run DocETL using Llama-70B or GPT-4o-mini for its optimizer so results (Table~\ref{tab:biodex}, \ref{tab:biodex_docetl}) are shown using GPT-4o for the DocETL optimizer. LOTUS supports both join and ranking. Although LOTUS provides a \verb|sem_topk| operator, we limit our ranking step to list-wise ranking to maintain a fair comparison with DocETL, and revisit ranking in Section~\ref{subsec:search}. 
Our main results are shown in Table~\ref{tab:biodex}, using Llama-70B and E5 embeddings and sampling size of $0.01\%$.
We also provide an additional set of benchmarks in Table~\ref{tab:biodex_docetl} comparing LOTUS and DocETL following the models, prompts and parameters used in the original DocETL preprint. The DocETL setup uses GPT-4o-mini, text-embedding-3-small embeddings and a sampling size of 500. Once again, DocETL fails when we attempt to use GPT-4o-mini for the optimizer, so we use GPT-4o. DocETL's join takes an LLM call budget, which we control to compare against the other baselines.

\begin{footnotesize}

\begin{table}[!t]

\centering
\caption{\small Biomedical Multi-label Classification Results on the BioDEX Dataset with Llama-70b }
\vspace{-.3cm}
\begin{threeparttable}
\begin{tabular}{p{3.0cm}p{.6cm}p{.7cm}p{1.4cm}p{.8cm}}
\toprule
Method & RP@5 & RP@10 & Execution Time (s) & \# LM Calls\\
\midrule
Search & 0.106 & 0.120 & 2.91 & 0.00 \\
AI UDF * & N/A & N/A & 2,144,560 & 6,092,500 \\
UQE & 0.115 & 0.114 & 6,559& 15,000 \\
DocETL Join (avg of 3 successes)**& 0.180& 0.219& 2050&  13,185\\
DocETL Join + Rank (avg of 3 successes)**& 0.262& 0.282& 2,342& 13,433\\
 \sys Join & 0.212 & 0.213 & 2,340  &5,644 \\
\sys Join + Rank  & 0.265& 0.280 & 2,503 & 5,869 \\
\bottomrule
\end{tabular}
\end{threeparttable}
\label{tab:biodex}
\end{table}

\end{footnotesize}

\begin{footnotesize}

\begin{table}[!t]

\centering
\caption{\small Additional Comparison Following DocETL's Original Benchmarks on the BioDEX Dataset with GPT-4o-mini} 
\vspace{-.3cm}
\begin{threeparttable}
\begin{tabular}{p{3.0cm}p{.6cm}p{.7cm}p{1.4cm}p{.8cm}}
\toprule
Method & RP@5 & RP@10 & Execution Time (s) & \# LM Calls\\
\midrule

DocETL Join + Rank (avg of 3 successes)**& 0.268& 0.287& 583&  35,669\\
\sys Join + Rank  & 0.289 & 0.296 & 647 & 32,026 \\

\bottomrule

\end{tabular}
\begin{tablenotes}
\centering
\footnotesize
\item[]* AI UDF is infeasible to run and latency is estimated under linear-scaling assumption in number of batched LLM calls.
\item[]** The DocETL baselines requires using GPT-4o as its optimizer and still exhibit frequent failures, with, on average, 1 in 3 runs failing. We rerun the system multiple times and report results averaged only over 3 successful runs. 
\end{tablenotes}
\end{threeparttable}
\label{tab:biodex_docetl}

\vspace{-.3cm}
\end{table}
\end{footnotesize}

\heading{Results.}
Table~\ref{tab:biodex} demonstrates the \sys programs consistently match or exceed the accuracy of all baselines, while also providing performance guarantees, unlike DocETL, which requires multiple reruns and manual selection of the optimizer LLM to produce high-quality results. The table reports the rank-precision@5 (RP@5) and rank-precision@10 (RP@10), following prior work~\cite{doosterlinck_-context_2024}, as well as execution time in seconds and the number of LLM calls. The \sys join with ranking program achieves higher rank-precision than the standalone \sys join program, and we show both.

We begin by informally comparing these accuracy results to that of D'Oosterlinck et al.~\cite{doosterlinck_-context_2024}, a state-of-the art AI pipeline that composed a multi-step DSPy~\cite{khattab_dspy_2023} program compiled using a combination of Llama-2-7b-chat, GPT-3.5-turbo, and GPT-4-turbo. D'Oosterlinck et al. report 24.73 RP@5 and 27.67 RP@10 for the compiled program, which is comparable to result quality achieved by the simple \sys join with ranking program.

Turning our attention to the baselines measured in Table~\ref{tab:biodex}, we see that, first, the search baseline offers low results quality, with the \sys programs achieving over $2\times$ higher RP@5 and RP@10. 
Next, The AI UDF baseline, using row-wise LLM calls, is prohibitively expensive, and reflects the cost of our gold algorithm for joins. Its quadratic scaling of LLM call complexity with respect to dataset size leads to $1,000\times$ higher estimated cost relative to the optimized LOTUS join program. UQE, which uses an embedding-based optimization, obtains similar accuracy to the search baseline, remaining well below the RP@5 and RP@10 of the LOTUS programs despite a generous LLM call budget, indicating that the UQE method has lower sample efficiency on this task. 
Next, we turn to a detailed comparison of DocETL and LOTUS, shown in Table~\ref{tab:biodex} and Table~\ref{tab:biodex_docetl}. 
LOTUS consistently matches or exceeds the accuracy of DocETL's successful runs. Notably, the DocETL agentic optimizer requires GPT-4o and frequently fails, requiring multiple reruns, while LOTUS' optimizer provides statistical accuracy guarantees, which we provide a detailed analysis on in Section~\ref{subsec:ablation}. Table~\ref{tab:biodex_docetl} also shows that the LOTUS and DocETL programs use a similar number of total LLM calls and execution time, however the token consumption per call of the DocETL programs are lower on average due to its optimized plan, which leads to modestly lower execution time despite modestly higher LLM calls.

\begin{footnotesize}

\begin{table}[!t]

\centering
\caption{\small Comparison of Candidate \sys Join Plans and Gold Algorithm on the BioDEX Dataset Using Llama-70b }
\vspace{-.3cm}
\begin{threeparttable}
\begin{tabular}{p{3.0cm}p{.6cm}p{.7cm}p{1.4cm}p{.8cm}}
\toprule
Method & RP@5 & RP@10 & Execution Time (s) & \# LM Calls\\

\hline
LOTUS Join Plan 1 & 0.1541 & 0.170 & 12,563 & 27,687 \\
LOTUS Join Plan 2 (chosen)  & 0.212 & 0.213 & 2,116& 5,290 \\
Gold Algorithm & N/A & N/A & 2,144,560 & 6,092,500 \\
\bottomrule

\end{tabular}
\end{threeparttable}
\label{tab:biodex_plans}

\end{table}

\end{footnotesize}

Lastly, we turn to Table~\ref{tab:biodex_plans} and study the candidate plans produced by the LOTUS join program. Plan 1, the sim-filter, requires more LLM calls to meet the recall and precision targets, compared to Plan 2, which is the project-sim-filter plan. This reflects that Plan 2's proxy offers a stronger signal for predicate evaluations. We see that the LOTUS optimizer selects Plan 2 to execute since it is lower cost, requiring fewer LLM calls. The selected pattern also results in significantly better accuracy due to its higher quality proxy signal for this task. Specifically, the project-sim-filter pattern offers $37\%$ higher RP@5 and $25\%$ higher RP@10 compared to the sim-filter, with $1,000\times$ fewer LLM calls than the gold algorithm. 

\subsection{Search \& Ranking}
\label{subsec:search}

\emph{Relevance-based ranking} has been widely studied in the context of information retrieval. In addition, our conversations with \sys users reveal a common need for ranking based on \emph{complex natural language criteria} (e.g., ranking customer reviews based on how frustrated they sound). In this application, we assess LOTUS' ranking capabilities on both types of ranking tasks with two datasets: BEIR's SciFact test set~\cite{thakur_beir_2021}, a widely used benchmark for retrieval and re-ranking, and a new dataset, HellaSwag-bench, which we generated to assess more complex ranking tasks. For both we report average nDCG@10, a standard ranking metric, and execution time (ET).
The SciFact dataset consists of scientific claims and a corpus of articles, where the task is to rank articles by relevance to a given claim. We sample 300 claims for our evaluation. HellaSwag-bench consists of 200 synthetic paper abstracts\footnote{Each paper abstract in HellaSwag-bench is synthetically generated by prompting Llama-70b to write a research abstract that claims a specified accuracy value, which we randomly sample from $0-100\%$.}, each reporting an accuracy on the HellaSwag~ dataset~\cite{zellers_hellaswag_2019}, and the task is to rank abstracts according to their reported accuracy. This dataset provides an objective ground truth, while focusing on a reasoning-based ranking criteria, rather than a relevance-based one\footnote{While the Hellaswag-bench ranking task might be efficiently solved with a sem\_map, to extract accuracy values, followed by a structured sorting, our evaluation focuses on assessing semantic ranking algorithms. We find this benchmark useful for understanding performance trade-offs, which may generalize to a wider set of reasoning-based ranking queries.}.
We report results for $n=20$ trials at temperature $t=0.7$, similar to prior works~\cite{khattab_dspy_2023}.

\heading{Baselines.} In addition to the standard search baseline, we add a re-ranking method using the MixedBread cross-encoder~\cite{noauthor_mixedbread-aimxbai-rerank-large-v1_2024}, a high quality re-ranker that is common in relevance-based retrieval. The AI UDF baseline performs a pointwise-ranking by prompting the LLM to assign a relevance score to each row, similar to prior work~\cite{zhuang_beyond_2024}. The DocETL baseline uses the system's LLM \verb|reduce| operation for ranking, as described by the pre-print~\cite{shankar_docetl_2024}.
Our \sys program uses the \verb|sem_topk| operator with a semantic index on the document corpus of both datasets. 
For the reranker, AI UDF, DocETL, and LOTUS programs on SciFact, we perform search to retrieve 100 articles, before re-ranking them with each method.

\begin{footnotesize}

\begin{table}[t]

\centering
\caption{\small Ranking Results on SciFact and Hellaswag-bench Datasets}
\vspace{-.3cm}
\begin{threeparttable}
\begin{tabular}{p{2.2cm}p{1cm}p{.6cm}p{.5cm}p{1cm}p{.6cm}p{.5cm}}
\toprule    
\multirow{2}{*}{Method} & \multicolumn{3}{c}{SciFact} & \multicolumn{3}{c}{HellaSwag-bench} \\
\cmidrule(lr){2-4} \cmidrule(lr){5-7}
& nDCG@10  & ET (s) & \# LM Calls & nDCG@10  & ET (s) & \# LM Calls\\
\midrule
Search &
0.712 & 0.009 & 0 & 
0.119 & 0.008 & 0\\

Reranker & 
0.741 & 2.64 & 0 &
0.461 & 2.36 & 0 \\

AI UDF &
0.457 & 9.83 & 100 &
0.091 & 19.7 & 200 \\

DocETL - Llama 70B optimizer (avg. of 3 sucesses)*&
N/A & N/A & N/A &
0.246 & 14.1 & 3 \\

DocETL - GPT 4o optimizer*&
N/A & N/A & N/A &
N/A & N/A & N/A \\






LOTUS &
0.765 & 36.3 & 213.2 &
0.919 & 57.0 & 506.4 \\

\bottomrule
\end{tabular}
\begin{tablenotes}
\centering
\footnotesize
\item[] * DocETL is unable to acheive any successful runs using GPT-4o for its optimizer on SciFact and HellaSwag-bench, nor using Llama-70B for its optimizer on SciFact
\end{tablenotes}
\vspace{-.3cm}
\end{threeparttable}
\label{tab:ranking_main}

\end{table}


\end{footnotesize}

\begin{footnotesize}

\begin{table}[t]
\centering
\caption{\small Comparison of LOTUS' Top-k to the Gold Algorithm and Other High-Quality Top-k Algorithms for Ranking}
\vspace{-.3cm}
\begin{threeparttable}
\begin{tabular}{p{2.2cm}p{1cm}p{.6cm}p{.5cm}p{1cm}p{.6cm}p{.5cm}}
\toprule    
\multirow{2}{*}{Method} & \multicolumn{3}{c}{SciFact} & \multicolumn{3}{c}{HellaSwag-bench} \\
\cmidrule(lr){2-4} \cmidrule(lr){5-7}
& nDCG@10  & ET (s) & \# LM Calls & nDCG@10  & ET (s) & \# LM Calls\\

\midrule 

Quadratic Topk  &
0.768 & 801.2 & 4950 &
0.886 & 2116.6 & 19,900 \\

Heap Topk  &
0.765 & 60.0 & 192.6 &
0.896 & 59.4 & 241.5 \\

Quickselect Topk  &
0.771 & 40.5 & 237.0 &
0.893 & 49.4 & 448.1 \\


LOTUS &
0.765 & 36.3 & 213.2 &
0.919 & 57.0 & 506.4 \\

\bottomrule
\end{tabular}

\vspace{-.3cm}
\end{threeparttable}
\label{tab:ranking_gold_algs}

\end{table}

\end{footnotesize}

\heading{Results.} 
We evaluated the performance of the \sys program against baselines, in comparison to the gold algorithm, and to alternative high-quality ranking algorithms we considered as gold algorithms for our \verb|sem_topk|. 

First Table~\ref{tab:ranking_main} demonstrates that LOTUS achieves $3\% - 100\%$ higher accuracy than the next best-performing baselines on Scifact and HellaSwag-bench. Specifically, as expected, the re-ranker offers competitive accuracy compared to the \sys program on Scifact, which focuses on relevance-based retrieval, the supervised task the re-ranker model was trained on. However, on HellaSwag-bench, the \sys program significantly outperforms the re-ranker on this reasoning-based task. We also observe that the point-wise AI UDF method and the list-wise ranking, used by DocETL, offer substantially lower accuracy, unable to outperform the re-ranking baseline. This reflects the limitations of point-wise and list-wise ranking, which has been studied in prior work~\cite{qin_large_2024}. Notably, the DocETL baseline altogether fails to produce any successful execution plans using the GPT-4o optimizer on Scifact and Hellaswag-bench and using the Llama-70B optimizer on Scifact.

Turning our attention to Table~\ref{tab:ranking_gold_algs}, we compare the quick-select top-k gold algorithm to several alternative candidates. The quadratic top-k, heap top-k, and quick-select top-k offer comparable accuracy, making each a viable gold algorithm. However, the three candidate choices provide significant differences in efficiency. 
The quadratic algorithm, which performs an LLM comparison between each pair of input documents, requires $20 - 82\times$ more LM calls and over $10\times$ higher execution time than the heap top-k and quick-select top-k.
In addition, the heap top-$k$ and quick-select top-$k$ methods offer interesting trade-offs in LLM call complexity and execution time. 
Notably the quick-select top-$k$ offers $16-32\%$ lower execution time than the heap-based sorting method, despite requiring more LLM calls in some cases. 
This is because the quick-select top-$k$ implementation allows for efficient batched processing in each round of the algorithm, whereas the heap-based top-$k$ incurs sequential LLM calls during heap updates. 
\sys currently implements the quick-select topk-$k$ by default, favoring lower execution time, although we future iterations may leverage multiple algorithms.

Finally, the Table~\ref{tab:ranking_gold_algs} compares the quick-select top-k with the \sys quick-select top-k using embedding-based pivot selection. We demonstrate that the optimization is lossless, and can decrease latency by $10\%s$. Specifically, this optimization is useful where document rank correlates with semantic similarity, which is likely on the Scifact dataset.

\begin{figure}[t]
  \centering
  \vspace{-.1cm}
    \centering
    \begin{footnotesize}
        \begin{tabular}{cc}
        \toprule
        ($\mu_1$) & Advancements in Recommender Systems and Multimodal Data Integration\\
        \hline
        ($\mu_2$) &Advancements in Generative Information Retrieval Systems\\                \hline

        ($\mu_3$) & Advancements in Large Language Models for Various Applications\\                \hline

        ($\mu_4$) & Advancements in AI Security and Malware Detection Techniques\\
        \hline

        ($\mu_5$) & Advancements in Robotic Navigation and Manipulation Techniques\\
        \bottomrule
        \label{tab:groupby}
        \end{tabular}
    \end{footnotesize}
    \vspace{-.5cm}
    \label{}
  \vspace{-.2cm}
  \caption{\small Discovered group labels from \sys \textit{sem\_group\_by("the topic of each \{paper\}", C=5)} over a dataset of recent ArXiv papers, scraped from the database (cs.DB), information retrieval (cs.IR), crytography and security (cs.CR) and robotics (cs.RO) domains.} 
  \label{fig:groupby_labels}
\vspace{-.5cm}
\end{figure}

\subsection{Topic Analysis on ArXiv Papers}
\label{subsec:arxiv_papers}
A common unsupervised discovery task requires grouping a large document corpus by topics and assigning descriptive labels to each group.
We consider this task over a dataset of 647 recent ArXiv articles, which we scraped from the database (cs.DB), information retrieval (cs.IR), cryptography and security (cs.CR), and robotics (cs.RO) domains. We aim to discover $5$ key topic labels.

We can succinctly represent this task using the \verb|sem_group_by| operator with LOTUS.
AI UDFs, DocETL and standalone search systems do not serve operations for unsupervised semantic-based group discovery. While UQE~\cite{dai_uqe_2024} studies a tractable optimization for serving non-LLM aggregations, e.g., COUNT, with LLM-based group-bys using stratified sampling, it does not focus on implementing, optimizing, and evaluating a standalone group-by operator, but suggests a standalone group-by implementation similar to our \verb|sem_group_by| gold algorithm. Thus, in this section, we focus our analysis on understanding the performance of our \verb|sem_group_by| gold algorithm and analyzing our optimization with respect to it. We use Llama-70B and E5 embeddings. For approximations, we use a sample size of 100 and $\delta=0.2$.

\heading{Results.}
The \verb|sem_group_by| consists of two sub-tasks: (a) discovering representative group labels, and (b) classifying each document to a discovered label. 

We first qualitatively analyze the labels discovered, in the first sub-task of the LOTUS implementation, which is equivalent to gold algorithm. As Figure~\ref{fig:groupby_labels} shows, the discovered labels intuitively align with recent research topics from the cs.DB, cs.IR, cs.CR, and cs.RO ArXiv domains related to recommendation systems, retrieval, LLM applications, AI security, and robotics. Performing this unsupervised group discovery took 44.03 seconds, representing a tractable implementation.

\begin{figure}[t]
  \centering
  \vspace{-.2cm}
    \includegraphics[width=0.5\linewidth]{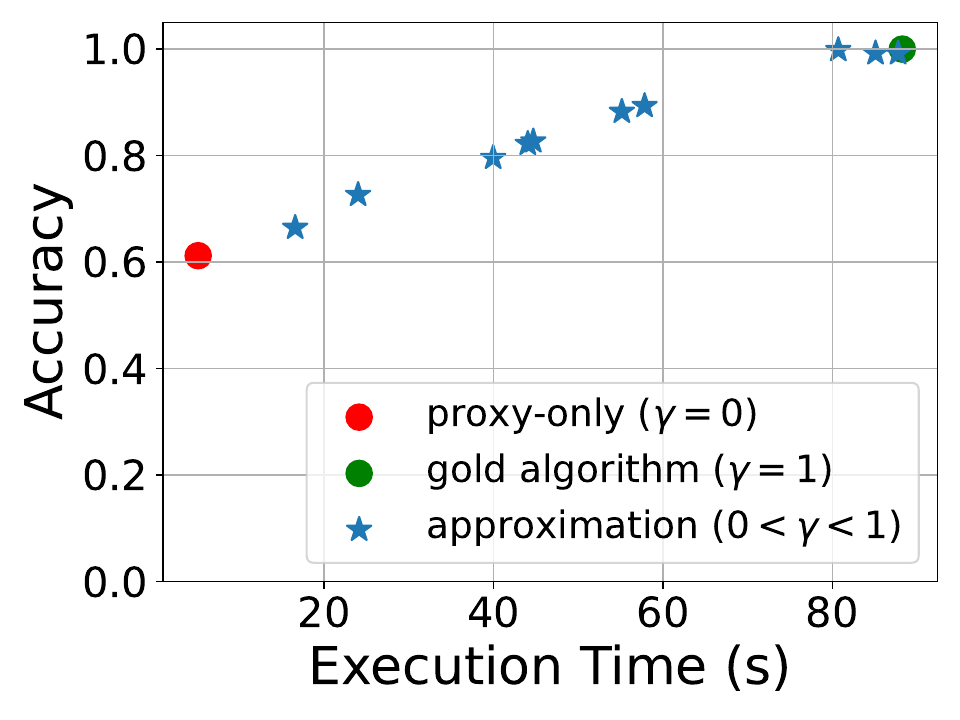}  
  \vspace{-.3cm}
   \caption{\small Classification accuracy vs. execution time (s) for the \sys sem\_group\_by over the ArXiv dataset. We compare the performance of implementing classifying each paper using the proxy only (red circle) or oracle only (green circle) to  our approximation with varied accuracy targets, at $\delta=0.2$ (stars).}
  \label{fig:groupby_cascades}
\vspace{-.3cm}
\end{figure}

Turning to the classification sub-task, we compare the performance of the gold algorithm to our approximation, which uses an embedding-based proxy here. Figure~\ref{fig:groupby_cascades} compares the execution time and classification accuracy of the oracle-based gold algorithm (green circle), our approximations using both the Llama-70B oracle LLM and embedding-based proxy at varied accuracy targets (blue stars), and the embedding-based proxy alone (red circle). We see that the proxy-only baseline is $17.4\times$ faster than the oracle, but about $39\%$ less accurate. Our approximation algorithm allows users to declaratively interpolate between these two extremes by varying the accuracy target. We also note that the sampling-based optimization procedure used to perform this optimization took less than 5 seconds, representing a small relative cost.

\begin{figure*}[t]
    \centering

    \includegraphics[width=0.6\linewidth]{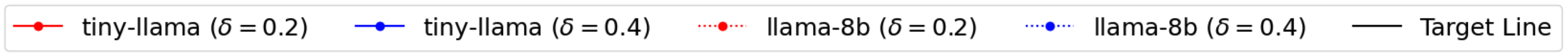}

    \begin{subfigure}[t]{0.2\linewidth}
        \centering
        \includegraphics[width=\linewidth]{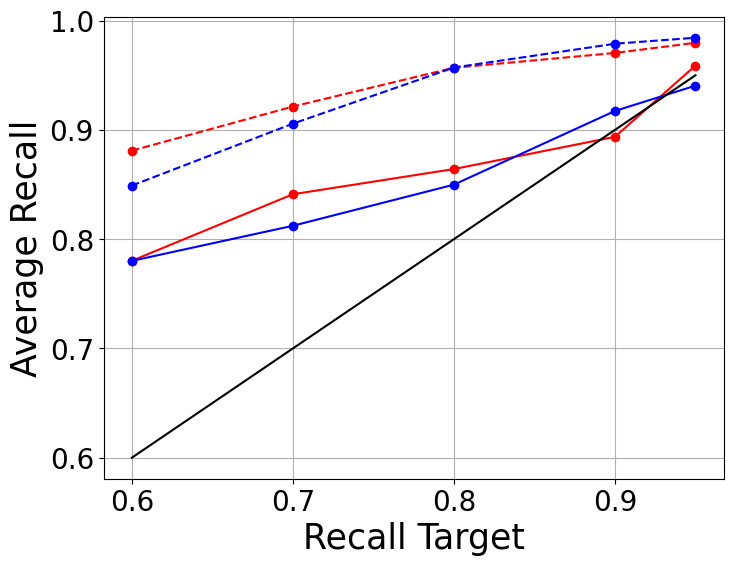}
        \caption{\small Observed vs Target Recall}
        \label{fig:recall_ablation}
    \end{subfigure}
    \hfill
    \begin{subfigure}[t]{0.2\linewidth}
        \centering
        \includegraphics[width=\linewidth]{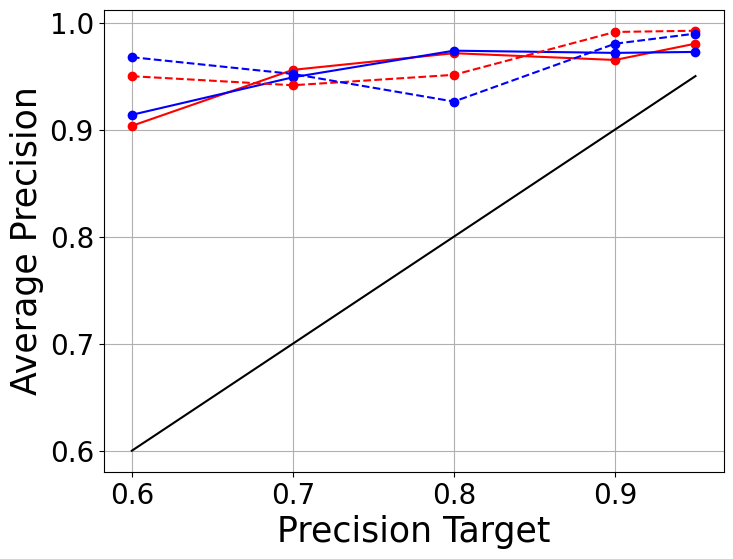}
        \caption{\small Observed vs Target Precision}
        \label{fig:precision_ablation}
    \end{subfigure}
    \hfill
    \begin{subfigure}[t]{0.2\linewidth}
        \centering
        \includegraphics[width=\linewidth]{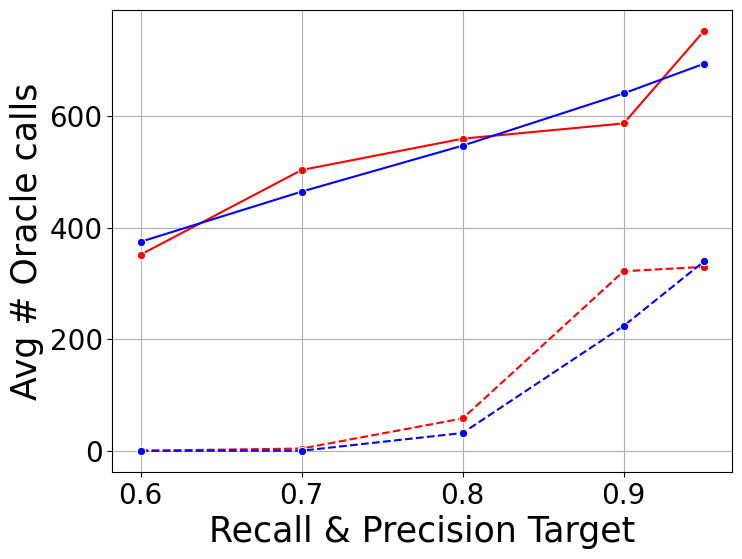}
        \caption{\small \# Oracle Calls vs Accuracy Target}
        \label{fig:numcall_ablation}
    \end{subfigure}
    \hfill
    \begin{subfigure}[t]{0.2\linewidth}
        \centering
        \includegraphics[width=\linewidth]{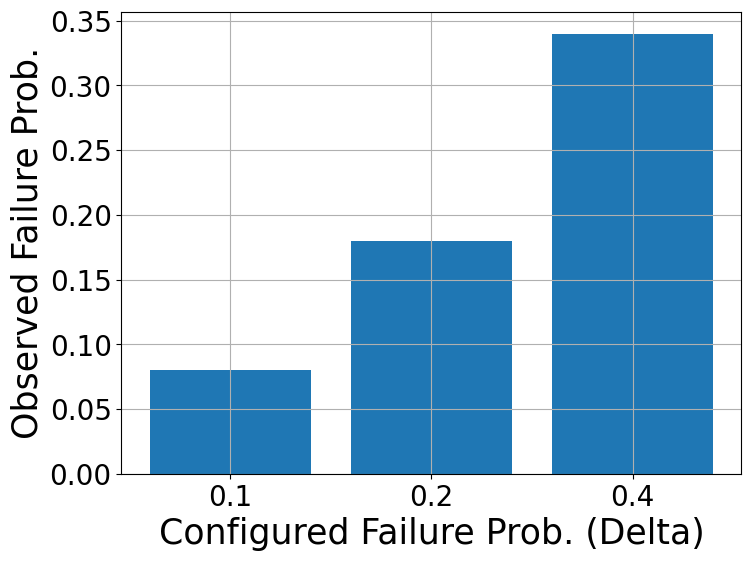}
        \caption{\small Observed vs Configured Failure Probability}
        \label{fig:delta_ablation}
    \end{subfigure}
    \vspace{-.3cm}
    \caption{\small We evaluate statistical accuracy guarantees with our semantic filter for fact-checking on the FEVER dataset. We show (a) average observed recall vs. configured recall targets ($\gamma_R)$, (b) average observed precision vs. configured precision targets ($\gamma_P)$, and (c) the number of oracle LLM calls vs. configured accuracy targets ($\gamma_R=\gamma_P$), using a Llama-70B oracle and either a Llama-8B proxy (dashed lines) or a weaker TinyLlama proxy (solid lines), for two different failure probabilities ($\delta=0.2, 0.4)$. We also show observed vs. configured failure probabilities using the TinyLlama proxy and Llama-70B oracle, for accuracy targets $\gamma_R = \gamma_P =0.9$.}
        \vspace{-.5cm}
    \label{fig:legend_and_subfigures}
\end{figure*}

\subsection{Accuracy Guarantees Evaluation}
\label{subsec:ablation}
Our approximation methods are designed to adaptively learn when to leverage a proxy to provide an accuracy guarantee, despite variations in proxy accuracy. To demonstrate the accuracy guarantees provided by our semantic operator optimizations, we study the \verb|sem_filter| performance on the fact-checking performance with varied proxy models, recall and precision targets and failure probability parameters. Figure~\ref{fig:recall_ablation} and Figure~\ref{fig:precision_ablation} demonstrate the average accuracy observed over 20 trials for varied recall targets and precision targets. We evaluate the filter approximation with both the Llama-8B proxy (solid lines) and a weaker proxy, with TinyLlama-1B (dashed lines) with a failure probability of $\delta=0.2$ (red) and $\delta=0.4$ (blue). As expected, the average recall and precision values increase with increasing accuracy targets for either metric. Figure~\ref{fig:numcall_ablation} demonstrates that decreasing the recall and precision target tends reduce the number of calls routed to the oracle. As expected, at a fixed accuracy target, our approximation requires more oracle calls when using the weaker proxy, in this case TinyLlama. Finally, in Figure~\ref{fig:delta_ablation} we record the percentage of 50 trials that do not meet the recall and precision target, both set to 0.9, using the TinyLlama proxy. The results confirm that the observed failure probabilities increase with the user-configured delta values, and are lower than the configured failure probabilities, as expected due to our conservative implementation.

\section{Related Work}

\heading{Data Systems for Row-wise LLM Operations.}
Many prior works focus on map-like, \emph{row-wise LLM-based operators}, either with generic AI user-defined functions (UDFs) or specialized operators ~\cite{liu_declarative_2024, lin_towards_2024, arora_language_2023, liu_suql_2024}. 
AI UDF systems~\cite{motherduck_introducing_nodate, williamdassafmsft_intelligent_2024, noauthor_snow_nodate,  databricks, noauthor_large_nodate, liu_optimizing_2024}  provide a low-level, non-declarative programming interface for row-wise LLM invocations. These systems also integrate vector search, equivalent to our \verb|sem_search| or \verb|sem_sim_join|, and support SQL queries that combine row-wise AI UDFs with non-LLM operations (e.g. average, count).
Alternatively, several recent works implement optimized row-wise LLM operations for \emph{specialized} tasks, including data cleaning, extract-transform-load (ETL) and conversational agents.
ZenDB~\cite{lin_towards_2024}, EVAPORATE ~\cite{arora_language_2023} and Palimpzest~\cite{liu_declarative_2024} provide LLM-based functions for extracting semi-structured documents into structured tables, logically similar to our \verb|sem_extract|, and Palimpzest also implements an LLM-powered \verb|filter|, logically similar to our \verb|sem_filter|.
SUQL~\cite{liu_suql_2024} augments SQL with a two logically row-wise operators, \verb|answer|, for questions-answering over each row, and \verb|summary|, for \emph{row-wise} LLM-based summarization, to serve knowledge-grounded \emph{conversational agents}.
In contrast to these works, semantic operators are a \emph{general-purpose query model} for semantic transformations over datasets, including both logically row-wise ones and more complex ones (e.g., joins, aggregations and ranking). 
Our detailed evaluation (Section~\ref{sec:eval}) demonstrates the performance limitations of  row-wise LLM execution models.

\heading{Best-Effort LLM-Based Analytics Systems.}
Several recent systems~\cite{anderson_design_2024, dai_uqe_2024, shankar_docetl_2024} go beyond row-wise LLM operators. However, these systems lack a formalism for LLM-based operators, providing no performance guarantees for their execution.
Following a preprint~\cite{patel_lotus_2024} of this work, DocETL~\cite{shankar_docetl_2024} proposes to use LLM agents as the query optimizer over LLM-powered operators. This is a promising direction for future work; however, our evaluation (Section~\ref{sec:eval}) demonstrates that DocETL's agentic approach leads to high variance in performance, no accuracy guarantees, and requires multiple reruns to achieve comparable accuracy to ours, with approximately 1 in 3 runs failing due to optimizer errors, indicating that more work is needed to make these methods robust.
UQE~\cite{dai_uqe_2024} also studies optimizations for LLM-powered operators, proposing an embedding-based approximation for LLM-powered filters. In contrast to our methods, which adaptively learn how to leverage a given proxy to meet accuracy targets, UQE's approach provides best-effort performance with no accuracy guarantees. UQE also studies optimizations for non-LLM aggregations (e.g., average, count) combined with LLM-based filter and group-by operators using stratified sampling, which is complementary to this work.

\heading{General ML-based Query Processing.}
Prior works study the use of non-LLM machine learning (ML) in databases. In contrast, our work integrates LLMs, which motivates our new formalisms for language-based operators and novel optimizations.
MADLib~\cite{hellerstein_madlib_2012} extends SQL with abstractions for descriptive statistics and machine learning (e.g., regression and classification). 
NoScope~\cite{kang_noscope_2017}, TASTI~\cite{kang_task-agnostic_nodate}, SUPG~\cite{kang_approximate_2020}, BlazeIt~\cite{kang_blazeit_2019} and probabilistic predicates~\cite{lu_accelerating_2018} propose methods to optimize queries involving expensive ML-based predicates (e.g., using object detectors) over large datasets, typically in video analytics. 
Some optimizations proposed in these works, such as model cascades and predicate re-ordering, are also useful for optimizing \sys pipelines with language models.

\heading{Table Question Answering.} 
A large body of work, including  retrieval-augmented generation~\cite{lewis_retrieval-augmented_2021} and Text2SQL~\cite{yu_typesql_2018, yaghmazadeh_sqlizer_2017, zhang_benchmarking_2024, zelle_1996-learning_nodate},  serves natural language interfaces for question-answering over databases. Typically, the LLM system invokes external tools, e.g., search APIs or SQL programs. Interestingly, these agentic workflows can instead invoke semantic operator programs as the underlying API for data processing. Recent work~\cite{biswal_text2sql_2024} demonstrates the promise of this approach to outperform baseline TableQA methods.

\heading{Adoption of Semantic Operators.} Based on a preprint~\cite{patel_lotus_2024} of this work, Google recently implemented experimental semantic operators in BigQuery Dataframes~\cite{noauthor_python-bigquery-dataframesnotebooksexperimentalsemantic_operatorsipynb_nodate}, including \verb|sem_map|, \verb|sem_filter|, \verb|sem_join|, \verb|sem_agg|, \verb|sem_topk|, \verb|sem_search|, and \verb|sem_sim_join|.

\section{Conclusion}
In this work, we proposed semantic operators for bulk semantic processing and provide the first formalized model for general-purpose, language-based transformations over data. 
Our results across a diverse set of applications, including fact-checking, biomedical multi-label classification, search, and topic analysis, demonstrate the generality, expressiveness and robustness of the semantic operator model as well as our optimization approach. For each task, we find that \sys programs capture state-of-the-art AI applications with low development overhead, match or exceed result quality of recent LLM-powered analytics systems, and substantially reduce cost, while ensuring accuracy guarantees. 
Since open-sourcing \sys, we have seen a growing user-base, as well as adoption of semantic operators among large database vendors. We believe both represent exciting steps towards new data systems that integrate powerful AI reasoning capabilities over vast knowledge corpora.
\begin{acks}
This research was supported in part by affiliate members and other supporters of the Stanford DAWN project, including Meta, Google, and VMware, as well as Cisco, SAP, and a Sloan Fellowship. Any opinions, findings, and conclusions or recommendations expressed in this material are those of the authors and do not necessarily reflect the views of the sponsors.
\end{acks}


\bibliographystyle{ACM-Reference-Format}
\bibliography{citations}

\end{document}